\documentclass[amsmath,amssymb,floatfix,superscriptaddress,twocolumn,prl]{revtex4}

\usepackage{graphicx}

\usepackage{amssymb}
\usepackage{amsmath}

\newcommand{\msub}[1]{_{\mathrm{#1}}}

\begin{document}


\title{Surface Encapsulation for Low-Loss Silicon Photonics}

\author{M.~Borselli}
\thanks{Present address: Xponent Photonics Inc., Monrovia, CA 91016}
\author{T.~J.~Johnson}
\affiliation{Thomas J. Watson, Sr., Laboratory of Applied Physics, California Institute of Technology, Pasadena, CA 91125}
\author{C.~P.~Michael}
\affiliation{Thomas J. Watson, Sr., Laboratory of Applied Physics, California Institute of Technology, Pasadena, CA 91125}
\author{M.~D.~Henry} 
\affiliation{Thomas J. Watson, Sr., Laboratory of Applied Physics, California Institute of Technology, Pasadena, CA 91125}
\author{O.~Painter}
\email{opainter@caltech.edu}
\affiliation{Thomas J. Watson, Sr., Laboratory of Applied Physics, California Institute of Technology, Pasadena, CA 91125}

\date{\today}

\begin{abstract} 
Encapsulation layers are explored for passivating the surfaces of silicon to reduce optical absorption in the 1500-nm wavelength band.  Surface-sensitive test structures consisting of microdisk resonators are fabricated for this purpose.  Based on previous work in silicon photovoltaics, coatings of SiN$_x$ and SiO$_2$ are applied under varying deposition and annealing conditions.  A short dry thermal oxidation followed by a long high-temperature N$_2$ anneal is found to be most effective at long-term encapsulation and reduction of interface absorption.  Minimization of the optical loss is attributed to simultaneous reduction in sub-bandgap silicon surface states and hydrogen in the capping material.
\end{abstract}


\maketitle

\noindent
The integration of high-speed photonics and electronics on a silicon platform is currently being actively pursued due to the soaring demand for bandwidth and the difficulties in maintaining Moore's Law scaling of microelectronics~\cite{ibm46-245,Pavesi-Si_photonics_2004}.  Due to the thermal and power limitations already faced by the electronics industry~\cite{IEEE-SMTMS-201,ibm50-339}, future systems incorporating on-chip optical processing and communication will likely put a heavy premium on low optical power levels and high optical efficiency.  With the many benefits of Si microphotonic circuits, such as the ability to create active and nonlinear optical elements, there is also the adverse effect of increased optical loss in comparison to conventional glass-based photonic circuits.  In addition to the intrinsic nonlinear absorption present in micron-scale Si waveguides~\cite{oe12-1611}, the high-index contrast of such structures results in significant optical scattering loss for surface-roughness at the nanometer-scale.  Over the past several years there has been considerable effort put forth to improve the quality of Si microphotonic device fabrication, with optical scattering loss having been reduced to a level of 1 dB/cm in single-mode waveguides~\cite{natpho1-65}.  This is still several orders of magnitude above the bulk absorption limit of moderately doped Si in the $1.3$--$1.5$ $\mu$m wavelength band~\cite{ieee-jqe23-123}, and it is interesting to consider whether further reductions in optical loss can be made.  One possible impedement is the absorption present at the Si surfaces of current high-index contrast and, consequently, high surface-to-volume ratio Si microphotonics.  

In previous work we measured the effects of surface chemistry on the optical losses in smoothly etched, high-$Q$ Si microdisk resonators~\cite{oe13-1515,apl88-131114} and found that for losses below the dB/cm level, surface absorption begins to play a role.  In that study, hydrogen-passivated surfaces showed marked improvement with optical losses measured as low as 0.15 dB/cm; however, the passivation was temporary and easily spoiled by the atmosphere.  In this Letter we investigate, for photonics applications, several techniques originally developed for electronic surface passivation of Si solar cells.  The reliability standards of Si photovoltaics fabrication~\cite{apl66-3636,sicaE14-11,sst16-167,ppra13-195} routinely call for various passivation layers to be deposited over the Si surfaces in order to preserve the lifetime of the minority-carrier electrons and holes.  Based upon our studies, we report a simple recipe consisting of a short dry thermal oxidation followed by a long high-temperature N$_2$ anneal that is effective at preserving the Si disks' high quality factors in the $1.5$-$\mu$m wavelength band, indicating that the Si/SiO$_2$ interface provides adequate dangling bond passivation for photonic applications.

Schmidt, \frenchspacing{et al.~\cite{sst16-167}}, found that plasma-enhanced chemical vapor deposition (PECVD) silicon nitride (SiN$_x$) passivation layers with or without an underlying thin thermal oxide layer achieved effective free-carrier lifetimes of $\sim$1\,ms, comparable to the best passivation schemes to date.  The optimized PECVD recipe for SiN$_x$ layers was found to result in nearly stoichiometric Si$_3$N$_4$ films and was generated through a direct parallel-plate reactor system.  Furthermore, they found that the effective lifetime further increased for the first 50\,min of a 400$^{\circ}$C anneal in forming gas ($5\%$ H$_2$, $95\%$ N$_2$) before eventually decaying with increased forming gas annealing (FGA).  As the effect of ambient hydrogen was found to be negligible, they attributed the increase of lifetime to the large reservoir of hydrogen in the SiN$_x$ ($\sim$15--20 at.\%) being released during the deposition and anneal.  The effective passivation was also found to benefit from an initial thermal oxide, providing a higher quality Si interface, before the SiN$_x$ deposition and concomitant diffusion of hydrogen through the SiO$_2$ to the Si surface.  In a more recent study, McCann \frenchspacing{et al.~\cite{ppra13-195}} found that a 25-nm thick thermal oxide grown at $900^{\circ}$C followed by a 30\,min FGA at $400^{\circ}$C was sufficient to obtain equally high effective lifetimes.  In addition, they found that the lifetimes could be spoiled by a 1-hr high-temperature 900--1000$^{\circ}$C N$_2$ anneal and subsequently repaired by another FGA.  They attributed this effect to reversibly adding or removing hydrogen from the Si/SiO$_2$ interface.

\begin{figure}

\centerline{\includegraphics[width=\columnwidth, bb = 40 30 520 220, clip = true]{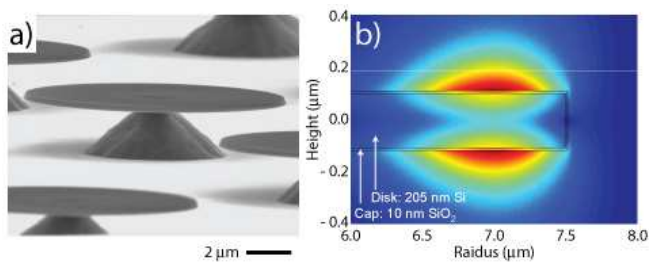}}
\caption{(a) An array of Si microdisk cavities with a radius of 5\,$\mu$m.  (b) Finite element simulation of the normalized electric-field intensity for the TM cavity mode studied in Fig.~\ref{fig:TOX2}.}
\label{fig:disk-n-field}
\end{figure}

We hypothesized that the effective electronic lifetime measurements from the previous references would be good indicators of optical loss at the Si interface.  Both PECVD SiN$_x$ and dry thermal oxide (TOX) layers were applied as encapsulation layers to Si microdisks (Fig.~\ref{fig:disk-n-field}).  The samples for these trials were fabricated from silicon-on-insulator (217-nm Si device layer, $p$-type 14--20\,$\Omega\cdot$cm, $\langle100\rangle$ orientation, 2-$\mu$m buried SiO$_2$ layer).  Microdisk resonators (with radii of 5, 7.5, and 10\,$\mu$m) were patterned with electron-beam lithography and a resist reflow process using ZEP520A resist and were defined with a SF$_6$-C$_4$F$_8$ inductively-coupled-plasma reactive-ion etch~\cite{oe13-1515}.  The devices were tested using an evanescent coupling technique~\cite{oe13-1515} employing a tunable laser source (1423--1496\,nm, linewidth $<$~300\,kHz) and a local fiber-taper probe~\cite{CM_dimple_taper}.  A fiber-based Mach-Zehnder interferometer was used to calibrate resonance linewidth $\delta\lambda$ measurements to $\pm$0.01\,pm---$\delta\lambda$ is related to the total cavity loss rate $\gamma$ by $\delta\lambda = \gamma\lambda\msub{o}^2/2\pi{}c$ where $\lambda\msub{o}$ is the free-space wavelength and $c$ is the speed of light.  

To ensure no ion contamination, all oxidations and anneals were done in a custom-built HF-cleaned quartz-tube oxidation furnace. Electronic grade II O$_2$, N$_2$, and forming gases were plumbed into the 2-inch diameter furnace using electro-polished stainless steel tubing.  Each gas's flow rate was independently controlled with a mass flow controller; typical flow rates inside the approximately 1-m long quartz furnace were $0.3$~standard liters per minute.  Thermal shock to the samples was avoided by using the unheated portion of the furnace tube as a second and intermediate staging area where the samples were allowed to slowly warm-up and cool-down before transfer to the heated portion of the chamber or being removed from the furnace.  To assess the loss of each surface treatment, the linewidth of a single surface-sensitive~\cite{apl88-131114} TM mode is tracked throughout a given processing sequence.   This procedure gives a long effective optical path length and ensures the surface is sampled in the same way after each step.

\begin{figure}
\centerline{\includegraphics[width=\columnwidth]{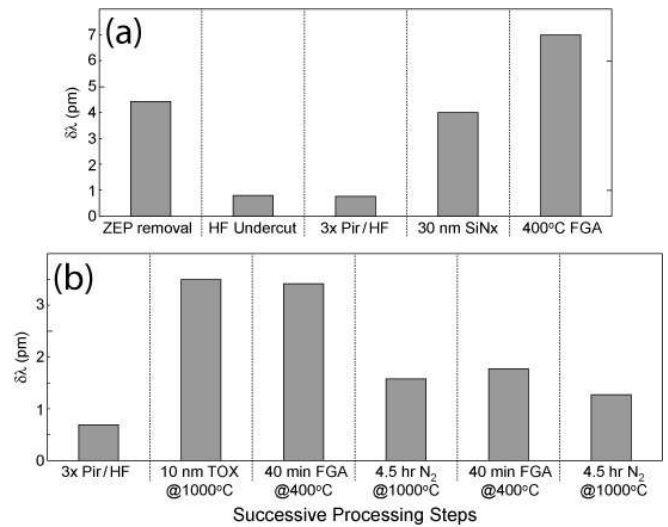}}
\caption{Summary of best linewidths after selected processing steps for 5--10\,$\mu$m radii disks fabricated with (a) a stoichiometric SiN$_x$ encapsulation layer and (b) a thermal oxide encapsulation layer along with various annealing trials.}
\label{fig:SiNx_TOX}
\end{figure}

The first encapsulation trial was done on a sample that had an initial 50-nm SiN$_x$ cap deposited on the surface prior to lithography and etching.  After removing the ZEP resist with an hour-long Piranha etch,  Fig.~\ref{fig:SiNx_TOX}(a) shows a bar graph summary of the best linewidths at selected intermediate points during the fabrication.  Having tested many devices on the sample, the best linewidth after the initial Piranha clean was $\delta\lambda=4.5$\,pm.  However, after an hour-long HF undercut which also removed the SiN$_x$ cap, the linewidths reduced considerably to $0.8$\,pm.  Further Piranha/HF treatments~\cite{apl88-131114} had no discernible effect on the linewidths.  

At this point, a nearly stoichiometric 30-nm thick SiN$_x$ encapsulation was deposited over the wafer.  Immediately prior to loading the wafer into the PECVD chamber, an additional 3$\times$ Piranha/HF treatment was done on the sample to ensure a clean and well-passivated Si surface.  The processing conditions were adapted from Ref.~\cite{sst16-167}, so a gas chemistry of 450\,sccm of 5\%\,SiH$_4$/N$_2$ and 50\,sccm of NH$_3$ was applied to the chamber held at 200\,mTorr and $400^{\circ}$C.  The gas was cracked with 60\,W from a 13.56-MHz radio-frequency source; no low frequency source was used in an attempt to suppress deposition damage by ion's oscillating below the $\sim$4\,MHz plasma frequency~\cite{photovoltaics1997-Lauinger}.  In contradiction to the photoconductance decay measurements of free-carrier lifetimes on Si solar cells~\cite{sst16-167}, the losses of the cavities significantly increased to 4\,pm, a difference of 3.2\,pm.  After testing with the ``as-deposited'' SiN$_x$ cap, the sample underwent a 40-min FGA at $400^{\circ}$C.  Subsequent testing revealed that the FGA had further deleterious effects on the samples, where the best linewidth was found to be 7\,pm as in Fig.~\ref{fig:SiNx_TOX}(a).  

Assuming a SiN$_x$ index of refraction of 1.9, finite element method simulations of the composite resonator show that TM modes possess $\Gamma_{\text{SiN}_x} = 11\%$ of the optical energy inside the SiN$_x$.  Thus, if the sources of loss were evenly distributed throughout the SiN$_x$, the material quality factor of the as-deposited material would be $Q_{\text{SiN}_x,\text{mat}}=\Gamma_{\text{SiN}_x}\lambda\msub{o}/ (\delta\lambda_{\text{after}} - \delta\lambda_{\text{before}}) \approx$ 5.0$\times$10$^4$, corresponding to an attenuation coefficient of $\alpha_{\text{SiN}_x,\text{mat}} \approx 7.1$\,dB/cm.  PECVD deposited material is known to have relatively high absorption coefficients (1--10\,dB/cm) due to \mbox{Si--H}, \mbox{O--H}, \mbox{N--H} bond absorption overtones in the telecommunications wavelength bands~\cite{jjap33-2593}.  Furthermore, as the FGA anneal would not be expected to harm the Si surfaces, a consistent interpretation of the results shown in Fig.~\ref{fig:SiNx_TOX}(a) is that any benefits of a hydrogenated Si surface were overwhelmed by the increased hydrogen content in the bulk SiN$_x$ layer.  This is also consistent with the fact that PECVD Si-rich nitride disks (with no high temperature anneal) were independently fabricated and tested achieving quality factors of 6$\times$10$^4$.    


A second sample underwent identical processing as the sample just described, including the deposition of a SiN$_x$ cap, lithography, dry-etching, ZEP removal, HF undercut, and $3\times$ Piranha/HF treatment.  As expected, the best measured linewidth of 0.7\,pm was very similar to the previous sample and is shown in Fig.~\ref{fig:SiNx_TOX}(b).  However, this time a 10-nm TOX layer was grown on the Si surface at $1000^{\circ}$C for 3.1\,min in an attempt to form a good Si interface with a hydrogen free material.  After switching off the O$_2$, the sample was allowed to cool slowly under an N$_2$ ambient for $\sim$5\,min before retesting.  The best linewidth after thermal oxidation was 3.5\,pm, a result similar to the SiN$_x$ cap.  However, the same 40-min FGA had virtually no impact on the sample with the thermal oxide cap.  A 4.5-hr high-temperature anneal in an N$_2$ ambient was found to significantly improve the losses, where the best linewidth was measured to be 1.6\,pm.  The high-temperature anneal consisted of holding the furnace at $1000^{\circ}$C for 3\,hr and then letting the temperature slowly ramp down to $400^{\circ}$C over the course of the remaining 1.5\,hr.  Assuming that the high-temperature anneal successfully healed the Si interface and bulk TOX, a 40-min FGA was conducted on the sample.  The FGA was found to slightly reduce the quality factor of the best resonance, where the linewidth was measured to be 1.8\,pm.  A second 4.5-hr high-temperature anneal, aimed at driving out any hydrogen added during the FGA step, successfully reduced the optical losses back down to $\delta\lambda=1.2$\,pm on the best resonance. This set of four anneals showed that the thermal oxide needed time at high temperature to remove material and surface defects.  Also, the FGA had a marginally degrading effect on the optical losses.

\begin{figure}
\centerline{\includegraphics[width=0.9\columnwidth]{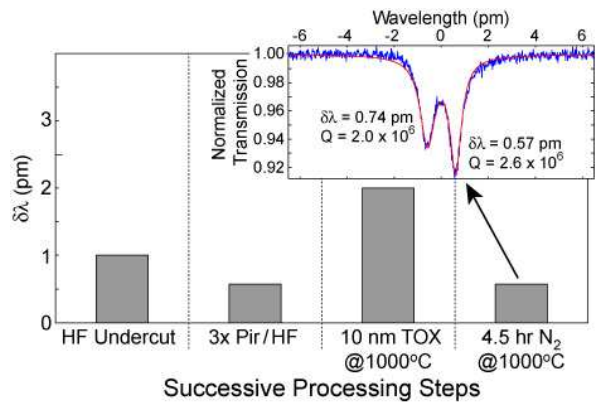}}
\caption{Summary of best linewidths after selected processing steps for 5--10\,$\mu$m radii disks fabricated without an initial protective cap.  Inset: transmission spectrum of a TM-like WGM doublet-resonance~\cite{oe13-1515} ($\lambda_{o}=1444.2$\,nm) of a 7.5-$\mu$m radius disk after the final high-temperature anneal.}
\label{fig:TOX2}
\end{figure}

A third sample was similarly fabricated but did not have an initial SiN$_x$ cap prior to lithography and dry-etching.  After HF undercutting and $3\times$ Piranha/HF treatments, the best linewidths were measured to be $1.0$ and $0.6$ pm, respectively (Fig. \ref{fig:TOX2}).  The marginal improvement in this case was attributed to a simplified single material dry-etch and a less damaged top Si surface.  The latter was confirmed after an identical 10-nm TOX cap with 5\,min cool-down showed a best linewidth of 2.0\,pm, much better than the second sample's 3.5\,pm linewidth after oxidation.  Omitting any FGA step, a final 4.5-hour high-temperature anneal healed the Si-interface and bulk silica cap, showing identical linewidths prior to oxidation.  Similarly processed samples in which the Si microdisk was completely oxidized through and then annealed had WGM resonances with $Q > 3${}$\times$10$^6$, indicating that the encapsulating oxide is of good optical quality after high-temperature anneals.  

Initial long term data show that microdisks with a TOX encapsulation layer and a high temperature N$_2$ anneal do not degrade after nine months stored in a clean-room environment ($\sim$22$^\circ$C, 44\% relative humidity). Going foward, formalized reliability testing under increased temperature, humidity, and pressure will still be necessary to prove true industrial applicability.  Furthermore, these tests could be coupled with materials analysis such as secondary ion mass spectroscopy (SIMS) or Auger profiling to yield even better insight into the physical mechanisms at work.  As a new vein of research, follow-up work should explore the sensitivity to furnace conditions including ramp-rate, gas chemistries, and anneal temperature.  Low-pressure chemical vapor deposition of SiN$_x$ could also be attempted to provide a high quality Si photonic passivation layer.    


The results shown in Fig.~\ref{fig:TOX2} represent the successful encapsulation of the once delicate Si-surfaces, as 10\,nm of thermal oxide will completely prevent native or chemical oxide growth during any subsequent fabrication steps.  While silicon nitride would have been slightly preferable in terms of chemical resistance, the high-quality thermal oxide is a simple and effective method of sealing the disks from environmental contamination while still allowing optical access to the mode's near-field.  Having also demonstrated this technology on a planar ring resonator~\cite{CM_dimple_taper}, this work represents a significant step towards developing ultra-low-loss silicon photonics technology.

{\frenchspacing This work was supported through the DARPA EPIC program.  For graduate fellowship support, we thank the Moore Foundation (MB and CPM), NPSC (MB), HRL Laboratories (MB), NSF (CPM), and the Hertz Foundation (MDH).}


\end{document}